\documentclass[aps,prd,twocolumn,fleqn,superscriptaddress]{revtex4}
\usepackage{graphicx,color,natbib}
\usepackage{amsmath,amssymb,amsfonts}

\newcommand{\bse}{\begin{subequations}}
\newcommand{\ese}{\end{subequations}}
\newcommand{\be}{\begin{equation}}
\newcommand{\ee}{\end{equation}}
\newcommand{\bea}{\begin{eqnarray}}
\newcommand{\eea}{\end{eqnarray}}
\newcommand{\ba}{\begin{array}}
\newcommand{\ea}{\end{array}}

\input amssym.def
\input amssym.tex

\usepackage[colorlinks=true, linkcolor=blue, bookmarks=true]{hyperref}

\begin{document}

\title{Entanglement Entropy in a Non-Conformal Background}
%

\author{M. Rahimi}
\email{{\rm{me}}$_{}$ rahimi@sbu.ac.ir}
\affiliation{Department of Physics, Shahid Beheshti University G.C., Evin, Tehran 19839, Iran}
\author{M. Ali-Akbari}
\email{{\rm{m}}$_{}$ aliakbari@sbu.ac.ir}
\affiliation{Department of Physics, Shahid Beheshti University G.C., Evin, Tehran 19839, Iran}
\author{M. Lezgi}
\email{mahsalezgee@yahoo.com}
\affiliation{Department of Physics, Shahid Beheshti University G.C., Evin, Tehran 19839, Iran}

\begin{abstract}
We use gauge-gravity duality to compute entanglement entropy in a non-conformal background with an energy scale $\Lambda$. At zero temperature, we observe that entanglement entropy decreases by raising $\Lambda$. However, at finite temperature, we realize that both $\frac{\Lambda}{T}$ and entanglement entropy rise together. Comparing entanglement entropy of the non-conformal theory, $S_{A(N)}$, and of its conformal theory at the $UV$ limit, $ S_{A(C)}$, reveals that  $S_{A(N)}$ can be larger or smaller than  $S_{A(C)}$, depending on the values of $\Lambda$ and $T$. 

\end{abstract}

\maketitle

%

\textit{Introduction:} The AdS/CFT correspondence states that type IIB string theory on the $AdS_5\times S^5$ background is dual to ${\cal{N}}=4$ $SU(N_c)$ superconformal gauge theory in a four-dimensional Minkowski space-time living on the boundary of the $AdS_5$ background \cite{ Maldacena, ahmad}. This outstanding correspondence is a strong-weak duality which makes it possible to investigate various strongly coupled systems. As a matter of fact, in the large number of colors and large t'Hooft coupling constant limit, the gauge theory is still a quantum theory but strongly coupled. However, the string theory reduces to a classical gravity which is a weakly coupled theory. Therefore, different questions in the strongly coupled gauge theory can be translated into corresponding problems in the classical gravity. This duality has been frequently applied to study various aspects of the strongly coupled systems such as quantum chromodynamics, quark-gluon plasma and condense matter, for instance see \cite{CasalderreySolana:2011us, Natsuume:2014sfa, Ali-Akbari:2015ooa}.

Since the AdS/CFT correspondence, or more generally gauge-gravity duality, applies to the non-conformal gauge theories as well as conformal ones, studying various effects of the non-conformal behaivor on the physical quantities is always an  attractive problem. A new family of solutions of a five-dimensional gravity model, including Einstein gravity coupled to a scalar field with a non-trivial potential, has been recently introduced and studied in \cite{Attems:2016ugt}. The corresponding four-dimensional strongly coupled gauge theory is not conformal and the theory has conformal fixed points at IR as well as at UV. this means that these solutions are asymptotic to the $AdS_5$ in the UV and IR limits with different radii. Different properties of the above background such as thermodynamics and relaxation channels have been studied \cite{Attems:2016ugt}. 

One interesting physical quantity, on the gauge theory side, is entanglement entropy \cite{Nishioka:2009un}. In the literature, gauge-gravity duality has been applied to investigate entanglement entropy successfully. For example, Entanglement entropy is also helpful to probe a confinement-deconfinement phase transition at zero temperature in confining theories \cite{Klebanov:2007ws}. Then search for transition has also been extended to non-conformal gauge theories at finite temperature \cite{Faraggi:2007fu}. It is shown that no transition takes place at finite temperature. In this paper we study the effect of introducing an energy scale on the entanglement entropy and to check the possibility of a phase transition in such a case \footnote{It is also important to notice that, in \cite{Pedraza:2014moa}, it is argued that entanglement entropy is more relevant timescale for the approach to equilibrium than two-point function and Wilson loop.}.

\textit{Model:} Here we review the non-conformal background introduced in \cite{Attems:2016ugt}. The background is a solution of five-dimensional gravity theory coupled to a scalar field with a non-trivial potential. The action of the gravity theory is given by %
\be\label{action} %
 S=\frac{2}{G_5^2}\int d^5x \sqrt{-g}\left(\frac{1}{4}{\cal{R}}-\frac{1}{2}(\nabla\phi)^2-V(\phi)\right),
\ee %
where $G_5$ is the five-dimensional Newton constant. The particular form of the potential is 
\be\label{potential}%
R_{UV}^2V=-3-\frac{3\phi^2}{2}-\frac{\phi^4}{3}+\left(\frac{1}{3\phi_M^{2}}+\frac{1}{2\phi_M^{4}} \right)\phi^6-\frac{\phi^8}{12\phi_M^{4}}.  
\ee %
An important point is that the potential has a maximum at $\phi=0$ and a minimum at $\phi=\phi_M$. It is shown that the resulting solution is asymptotically $AdS_5$ in the UV($\phi=0$) limit with radius $R_{UV}$. Moreover the solution near $\phi=\phi_M$, in the IR limit, approaches $AdS_5$ as well with a different radius $R_{IR}$. The relation between the radii of the $AdS_5$ backgrounds is given by%
\be %
 R_{\rm{IR}}=\frac{1}{1+\frac{\phi_M^2}{6}}R_{\rm{UV}},
\ee %
which clearly indicates that $ R_{\rm{IR}}< R_{\rm{UV}}$. According to gauge-gravity duality, the number of degrees of freedom in the gauge theory is related to the radius of the background. Thus a smaller number of degrees of freedom lives in the IR limit. 

The vacuum solution for arbitrary $\phi_M$ can be analytically expressed in the form 
\be%
 ds^2=e^{2A(r)}\left( -dt^2+dx^2\right)+dr^2 ,
\ee %
where 
\be\begin{split}%
 e^{2A}&=\frac{\phi_0^{2}}{\phi^2}\left(1-\frac{\phi^2}{\phi_M^{2}} \right)^{\frac{\phi_M^{2}}{6}+1}e^{-\frac{\phi^2}   {6}}, \cr 
 \phi(r)&=\frac{\phi_0e^\frac{-r}{R_{uv}}}{\sqrt{1+\frac{\phi_0^{2}}{\phi_M^{2}}e^\frac{-2r}{R_{uv}}}}.
\end{split}\ee %
$\phi_0$ is a constant corresponding to the source $\Lambda$ of the scalar operator on the gauge theory side. It is also related to an energy scale $\Lambda$ via $\Lambda=\phi_0/R_{UV}$. After two successive change of coordinates as follows.
\be\begin{split} %
u&=e^{-r/R_{uv}},\cr
z(u)&=\int_{0}^{u} du\frac{R_{UV}}{u}e^{-A},
\end{split}\ee 
we finally obtain
\be\label{metric}%
 ds^2=\frac{R_{\rm{eff}} (z)^{2}}{z^2}\left(-dt^2+dx^2+dz^2 \right), 
\ee %
where $R_{\rm{eff}}(z)=z e^{A}$. Parameter $R_{\rm{eff}}(z)$ varies between the radii of the $AdS_5$ backgrounds in the UV and IR limits corresponding to $R_{\rm{UV}}$ and $R_{\rm{IR}}$, respectively.

At finite temperature, the solution in the Eddington-Finkelstein coordinate is 
\be 
ds^2=e^{2A}\left(-h(\phi)d\tau^2+dx^2 \right)-2e^{A+B} R_{UV} d\tau d\phi,
\ee
with $h(\phi)$ vanishing at the horizon, i.e. $h(\phi_h)=0$. Solving Einstein's equations obtained from \eqref{action}, the different metric components are given by  \cite{Attems:2016ugt}
\bse\begin{align}%
A(\phi)&=-\log\left(\frac{\phi}{\phi_0}\right)+\int_{0}^{\phi}d\tilde{\phi}\left(G(\tilde{\phi})+\frac{1}{\tilde{\phi}} \right),  \\
B(\phi)&=\log\left(|G(\phi)|\right) +\int_{0}^{\phi}d\tilde{\phi}\frac{2}{3G(\tilde{\phi})},\\
h(\phi)&=-\frac{e^{2B(\phi)}L^2\left(4V(\phi)+3G(\phi)V'(\phi) \right) }{3G'(\phi)}.
\end{align}\ese %
The function $G$ must satisfy the following equation 
\be\begin{split}%
\frac{G'(\phi)}{G(\phi)+\frac{4V(\phi)}{3V'(\phi)}}&=\frac{d}{d\phi}\log\bigg[\frac{1}{3G(\phi)}-2G(\phi) \cr
&+\frac{G'(\phi)}{2G(\phi)}-\frac{G'(\phi)}{2\left(G(\phi)+\frac{4V(\phi)}{3V'(\phi)}\right)}\bigg], 
\end{split}\ee %
where, using the Einstein's equations, its behaviour near horizon is
\be\begin{split}
G(\phi)&=-\frac{4V(\phi_H)}{3V'(\phi_H)} \cr
&+\frac{2}{3}\left(\phi-\phi_H \right)
\left(\frac{V(\phi_H)V''(\phi_H)}{V'(\phi_H)^2}-1\right),
\end{split}\ee  %
up to second order in $\phi-\phi_H$. Computing the Hawking temperature of the above solution yields
\be\label{Tlambda} %
\frac{T}{\Lambda}=-\frac{R_{UV}^2V(\phi_H)}{3\pi\phi_H}\exp\{\int_{0}^{\phi_H}d\phi\left(G(\phi)+\frac{1}{\phi}+\frac{2}{3G(\phi)}\right)\}.
\ee

Based on numerical results \cite{Attems:2016ugt}, at high temperature ($T\gg\Lambda$), the gauge theory behaves as a conformal theory although the trace of the stress tensor is not zero. In the opposite limit, at low temperature ($T\ll\Lambda$), the gauge theory is conformal as well. Furthermore, it is clearly seen that $\phi_M$ evaluates the non-conformality of the theory or in the other words the larger $\phi_M$ the larger deviation from conformality. Finally a significant result is time ordering of relaxation times. In fact the system under consideration can either be firstly isotropised, meaning that all pressures become equal or be firstly equilibrated, meaning that the equation of state becomes applicable. Quasi-normal mode calculation indicates that at high temperature the system first isotropises and subsequently equilibrates. At low temperature the time ordering is reversed. For more details, we refer the reader to the original paper.

Before closing this part, we would like to emphasize that various time scales of relaxation for an out-of-equilibrium system has been firstly studied in \cite{Ali-Akbari:2016sms}.

\textit{Entanglement Entropy:} Let us consider a quantum system with many degrees of freedom at zero temperature which is described by a pure ground state $| \psi \rangle$. Therefore the density matrix is given by $\rho=| \psi \rangle\langle \psi |$. One can divide the mentioned system into two subsystems $A$ and $B$. The observer who is restricted to live in the subsystem $A$ does not have access to the degrees of freedom of subsystem $B$. Thus its density matrix can be found by taking the trace over these degrees of freedom, i.e. $\rho_A=tr_B\rho$. Then entanglement entropy of the subsystem $A$ is defined as $S_A=-tr_A(\rho_A\log\rho_A$). This quantity states how much information is lost when an observer is restricted to the subsystem $A$. For a gauge theory in $d>2$ space-time, dimension the leading divergence of $S_A$  is proportional to the area of the subsystem $A$. For a two-dimensional conformal gauge theory, where the subsystem $A$ is an interval of length $l$, the entanglement entropy can be analytically calculated as a universal result $S_l=\frac{c}{3}\log(\frac{l}{a})$ where $c$ and $a$ are central charge and the UV cut-off of the field theory, respectively. 

On the holographic side, entanglement entropy calculation has a simple prescription. It is proposed that entanglement entropy $S_A$ can be computed from the following formula
\be %
 S_A=\frac{\rm{Area}(\gamma_A)}{4G_5},
\ee %
where $\gamma_A$ is a three-dimensional minimal area surface in asymptomatically $AdS_5$ background whose boundary is given by $\partial A$ (which is the boundary of the subsystem $A$). This prescription perfectly produces well-known results, such as entanglement entropy in two-dimensional conformal field theory, and therefore it is reliable to compute the entanglement entropy in the strongly coupled gauge theories. For more details see for example \cite{Nishioka:2009un, Srednicki:1993im, Harlow:2014yka} and references therein.

\textit{Numerical results:} We start with a general form for the background as
\be\label{metric1}
 ds^2=-f_1(z)dt^2+f_2(z) dz^2+f_3(z)d\vec{x}^2 ,
\ee 
where $z$ is the radial direction. The background is asymptotically $AdS_5$ and its boundary is located at $z=0$. In order to determine entanglement entropy, we have to divide the boundary region into two subsystems $A$ and $B$. Subsystem $B$ is defined by $-\frac{l}{2}<x_1(\equiv x)<\frac{l}{2}$ and $x_2, x_3\in(-\infty,+\infty)$ at a given time. Then the minimal area of $\gamma_A$, which is proportional to entanglement entropy of subsystem $A$, is obtained by minimizing the following area 
\be\label{area} %
 S_A=\frac{1}{4G_5}\int d^3 x \sqrt{g_{in}},
\ee %
where $g_{in}$ is induced metric on $\gamma_A$. Using \eqref{metric1}, it is easy to see that  
\be\label{EE1}
S_A=\frac{V_2}{4G_5}\int_{\frac{-l}{2}}^{\frac{l}{2}} dx \sqrt{f_3^3(z)+f_3^2(z) f_2(z) z^{\prime 2}
},
\ee
where $V_2$ is the area of two-dimensional surface of $x_2$ and $x_3$ and $z'=\frac{dz}{dx}$. The above area does not depend on $x$ explicitly and thus the corresponding Hamiltonian is a constant of motion
\begin{align}\label{derevative}
\frac{f_3^2\left(z\right)}{\sqrt{f_3\left(z\right)+f_2\left(z\right)z^{\prime 2}}}={\rm{const}}=f_3^\frac{3}{2}(z_*),
\end{align}
where $z_*$ is the minimal value of $z$, i.e. $z(x=0)=z_*$, and $z'(x=0)=0$. Hence, from \eqref{derevative}, we find 
\begin{align}\label{zprime}
z^\prime=\sqrt{\frac{f_3(z)}{f_2(z)}}\sqrt{\frac{f_3^3(z)}{f^3_3(z_*)}-1},
\end{align}
and then the relation between $l$ and $z_*$ can be easily obtained 
\begin{align}
l=2\int_0^{z_*}\sqrt{\frac{f_2(z)}{f_3(z)}}\frac{dz}{\sqrt{\frac{f_3^3(z)}{f_3^3(z_*)}-1}}.
\end{align}
Finally, by substituting \eqref{zprime} in \eqref{EE1}, we have 
\be\label{entanglement}
S_A=\frac{V_2}{2G_5}\int_{0}^{z_*}\frac{f_3^\frac{5}{2}\left(z\right)f_2^{\frac{1}{2}}\left(z\right)}{\sqrt{f_3^3\left(z\right)-f_3^3\left(z_*\right)
}}dz.
\ee
As it is obvious the factor behind the integral in \eqref{entanglement} is a constant and as a result we only need to compute the integral which is proportional to the entanglement entropy. 

\begin{figure}
\begin{center}
\includegraphics[width=70mm]{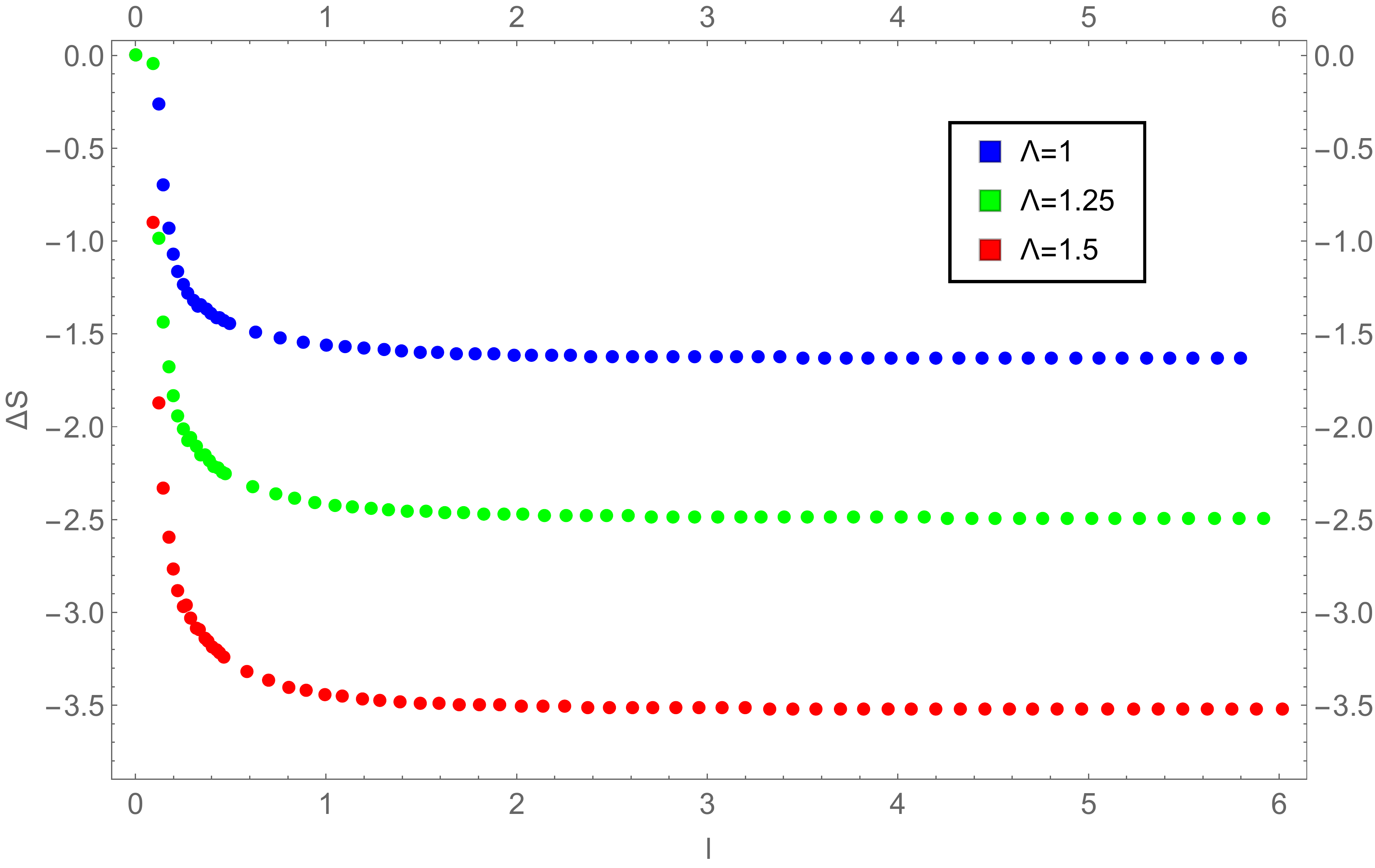}
\caption{$\Delta S$ for three different energy scales as a function of $l$ with $\phi_M=100$.
}\label{eel}
\end{center}
\end{figure}

At zero temperature, according to \eqref{metric}, it is clear that $f_1(z)=f_2(z)=f_3(z)=\frac{R_{eff}^2(z)}{z^2}$. In figure \ref{eel}, we have plotted the difference of the entanglement entropies for non-conformal and conformal(in the UV limit) field theories, i.e. $\Delta S\equiv G_5\Delta S=S_{A(N)}-S_{A(C)}$, as a function of $l$. Since we would like that a sizeable non-conformality emerge in the field theory, the value of $\phi_M$ is chosen to be large. As usual, the non-conformal field theory becomes conformal in the UV limit, which is probed by very small $l$, and therefore $\Delta S$ must go to zero in this limit. In other words, on the gravity side, $z_*$ is close to the boundary for very small values of $l$. Therefore, near boundary region, which is $AdS_5$ with good accuracy, contributes more to $S_{A(N)}$. By increasing $l$, $z_*$ reaches deeper in the bulk and thus the deviation of geometry from $AdS_5$ can be realized further by minimal area or equivalently entanglement entropy. On the field theory side, since in our model the number of degrees of freedom decreases from boundary to the bulk and considering the point that entanglement entropy is proportional to the number of degrees of freedom, for a given value of $l$ the non-conformal entanglement entropy $S_{A(N)}$ is smaller in comparison to its value in the UV limit, $S_{A(C)}$. As a result, $\Delta S$ is always negative at zero temperature.  For large enough values of $l$, $z_*$ approaches $AdS_5$ in the IR limit with smaller radius $R_{IR}$, meaning that the number of degrees of freedom becomes constant and proportional to $R_{IR}$. Therefore the difference between the number of degrees of freedom in the UV and IR limit becomes constant and accordingly $\Delta S$ goes to a constant value.  The above discussion is confirmed by figure \ref{eel}. Furthermore, by increasing energy scale $\Lambda$, the entanglement entropy of the non-conformal theory decreases. It is also significant to note that $\Delta S$ does not change for large enough values of $l$. 
\begin{figure}
\begin{center}
\includegraphics[width=70mm]{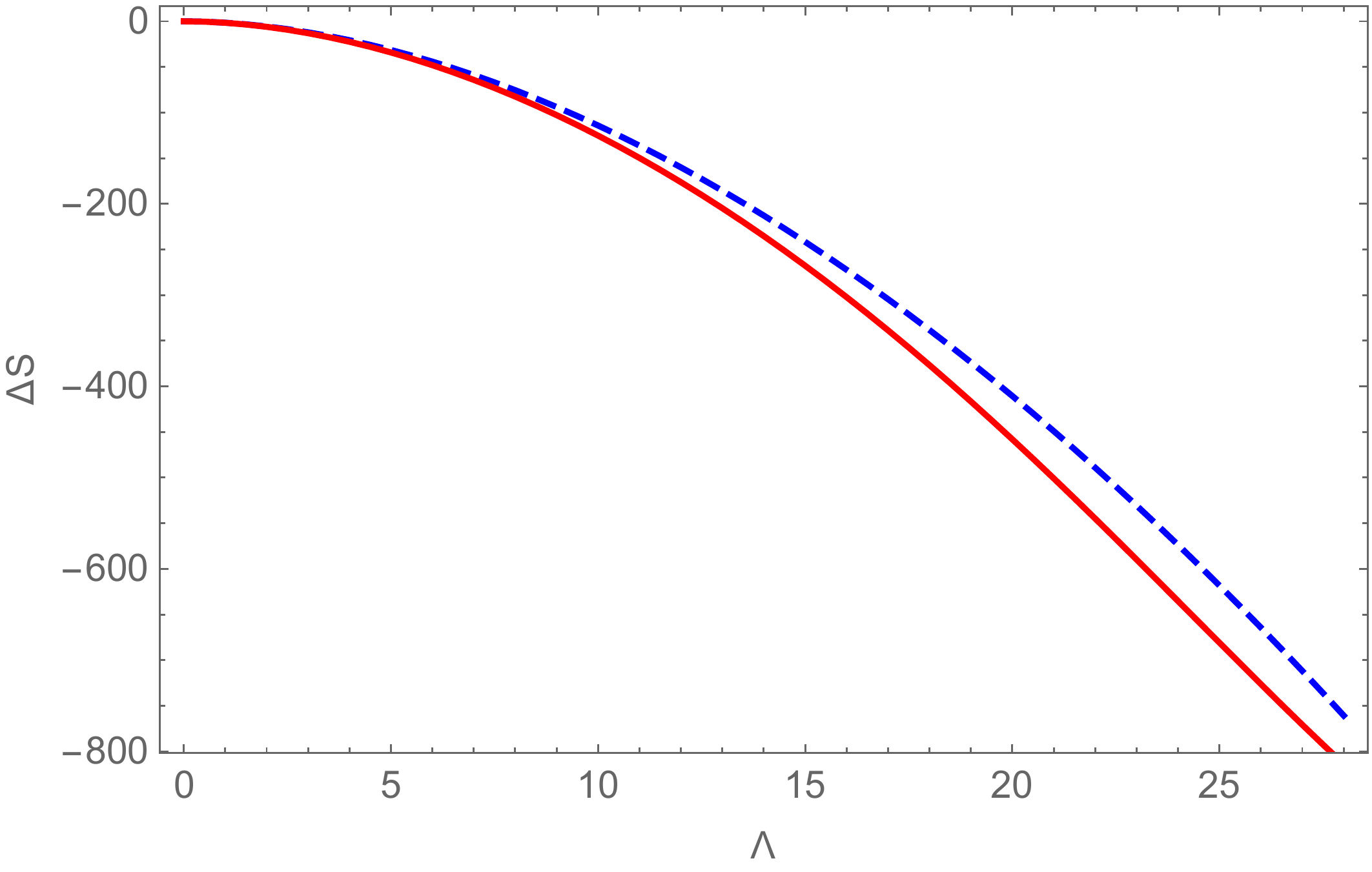}
\caption{$\Delta S$ in terms of energy scale $\Lambda$ for $\phi_M=1$(Blue) and $\phi_M=100$(Red) and $l=2.2$.}\label{deltaSlambda}
\end{center}
\end{figure}

\begin{figure}
\begin{center}
\includegraphics[width=70mm]{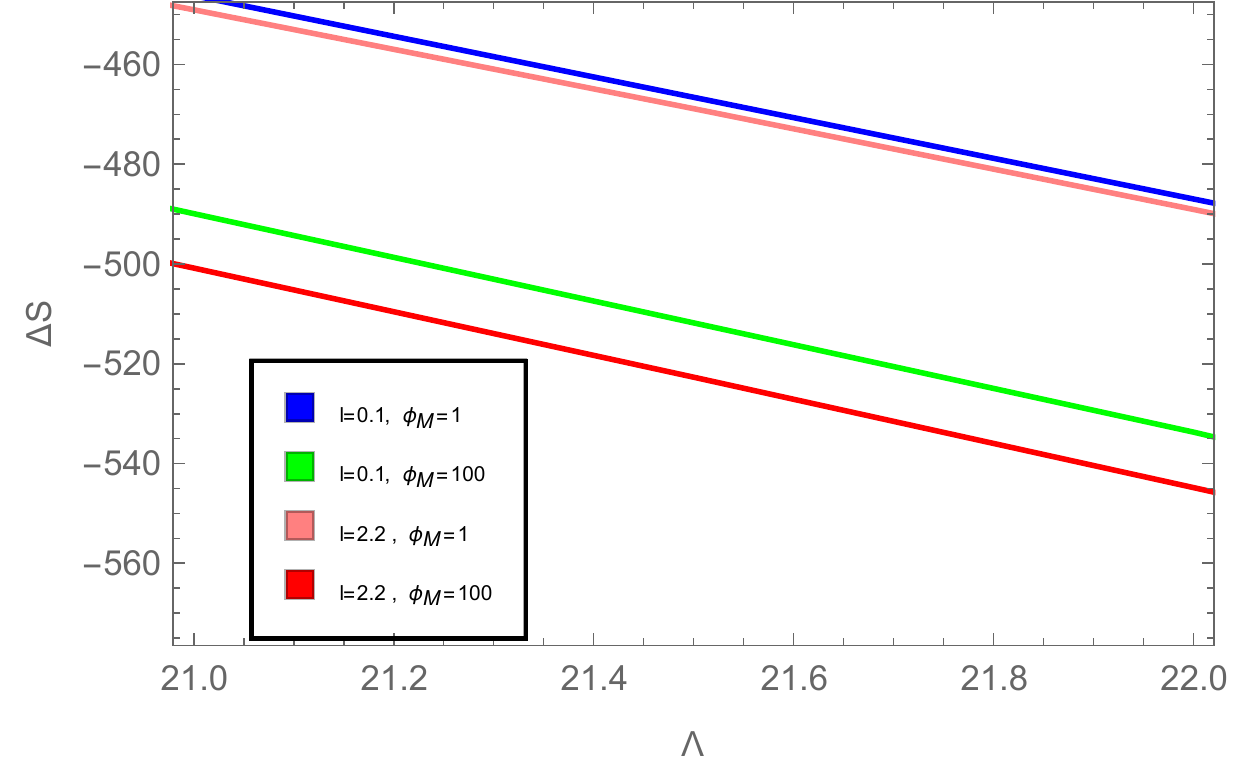}
\caption{$\Delta S$ in terms of energy scale $\Lambda$.
}\label{fignew2}
\end{center}
\end{figure}

The mere introduction of an energy scale $\Lambda$ breaks conformal symmetry in the model and the non-conformality is controlled by $\phi_M$. In figure \ref{deltaSlambda}, for given $l$, we have plotted $\Delta S$ in terms of $\Lambda$ for two values of $\phi_M$. This figure shows that non-conformal entanglement entropy decreases by raising $\Lambda$, in agreement with figure \ref{eel}. For small enough values of energy scale, $\Delta S$ is almost independent of non-conformality. However, for larger values of $\Lambda$ there is a decrease in the non-conformal entanglement entropy due to non-conformality. Note that there is a slight difference between the two curves at large $\Lambda$. This behaviour is confirmed for different values of $l$ by our numerical results. In particular, in figure \ref{fignew2}, $\Delta S$ is plotted in the range of $\Lambda= 21-22$. At fixed $\Lambda$, $\Delta S$ is smaller for the case with $l=0.1$, as expected.

\begin{figure}
\begin{center}
\includegraphics[width=80mm]{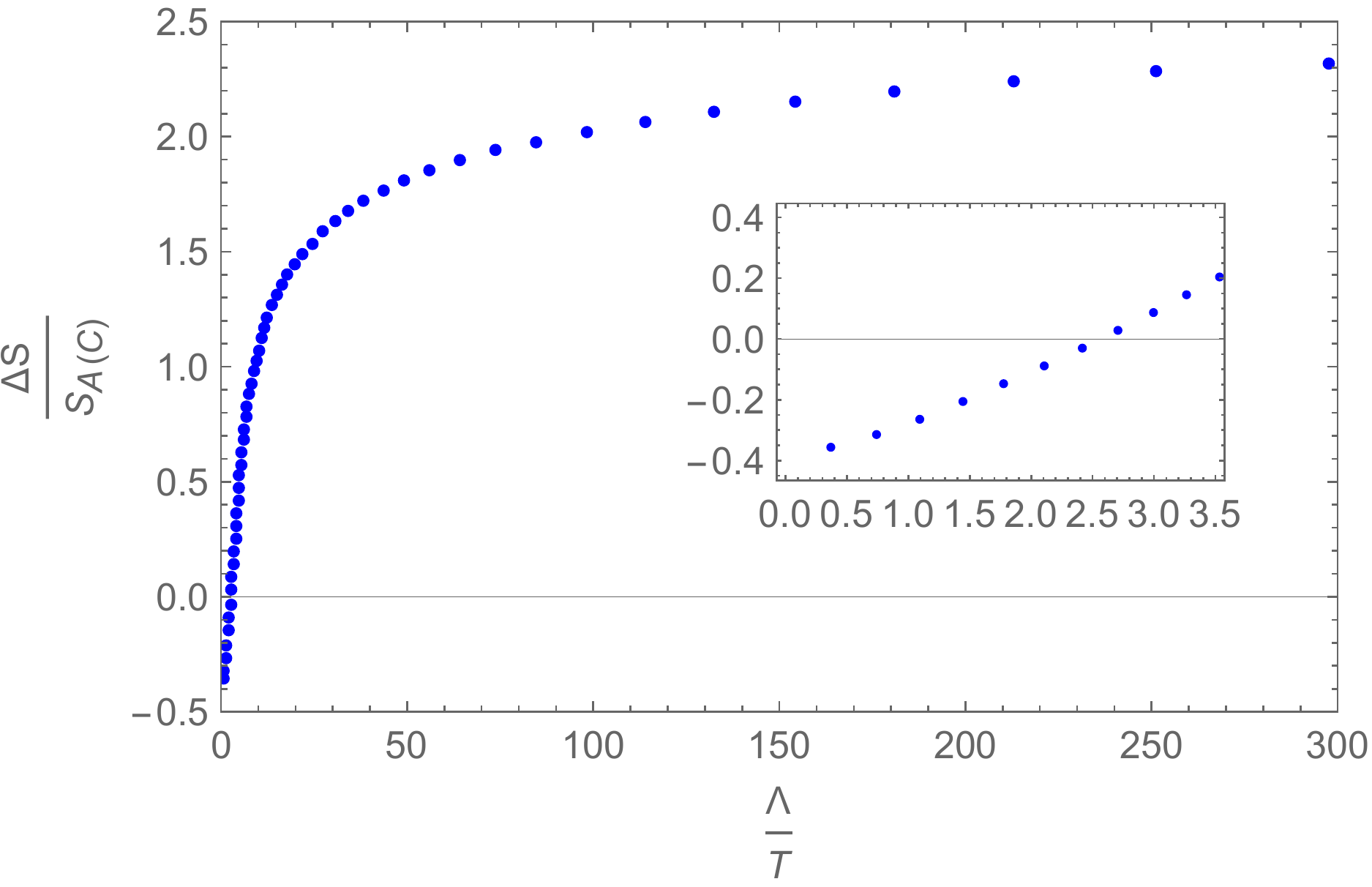}
\caption{$\Delta S=S_A(\Lambda=0.8,\phi_H\geq 0.1)-S_{A(C)}(\phi_H=0.1)$ as a function of $\frac{\Lambda}{T}$ for $\phi_M=100$ and $l=0.21$.}\label{differenT}
\end{center}
\end{figure}
\begin{figure}
\begin{center}
\includegraphics[width=75mm]{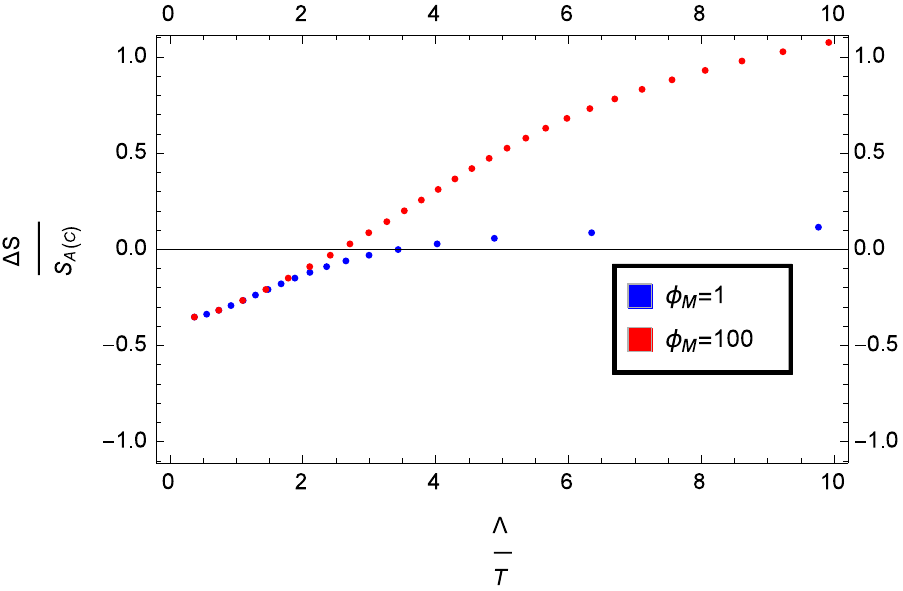}
\caption{$\Delta S=S_A(\Lambda=0.8,\phi_H\geq 0.1)-S_{A(C)}(\phi_H=0.1)$ as a function of $\frac{\Lambda}{T}$ for $l=0.21$.}\label{differenM}
\end{center}
\end{figure}

At finite temperature, we have 
\be\label{met}\begin{split}
f_2(\phi)&=\frac{R_{UV}^2}{h\left(\phi\right)}e^{2B\left(\phi\right)},\cr
f_3(\phi)&=e^{2A\left(\phi\right)}.
\end{split}\ee
Using \eqref{Tlambda}, \eqref{entanglement} and \eqref{met}, $\frac{\Delta S}{S_{A(c)}}$ is plotted as a function of $\frac{\Lambda}{T}$ for given $l$. We, first, focus on $\frac{\Lambda}{T}\ll 1$.  By carrying out a simple numerical analysis, it turns out that $\frac{\Lambda}{T}$ goes to zero when $\phi_H\rightarrow 0$. As a result, for very small values of $\frac{\Lambda}{T}$ the horizon is close to the boundary which makes it difficult to find a reliable numerical solution. But there is an intuitive explanation for this high temperature region. Since in the high temperature limit the non-conformal background approaches $AdS_5$ with radius $R_{UV}$, we expect that $\Delta S\rightarrow 0$, similar to the zero temperature case. Now let us consider a small enough value for $T$ in such a way that $\Lambda\ll T$. Taking these assumptions, the case at hand and zero temperature case are nearly alike and hence one expects that $\Delta S$ increase from negative values to zero by decreasing $\Lambda$, as it is clearly seen in figure \ref{deltaSlambda}.

Apart from the mentioned region in the above paragraph, our numerical results are shown in figure \ref{differenT}. By varying $\frac{\Lambda}{T}$ we observe that $\frac{\Delta S}{S_{A(c)}}$ changes sign at a point denoted by $(\frac{\Lambda}{T})_*$. For $\frac{\Lambda}{T}<(\frac{\Lambda}{T})_*$,  $\Delta S$ is always negative and therefore $S_{A(N)}$ is smaller than $S_{A(C)}$. The discussion for negative values of $\Delta S$ is similar to the case with $T=0$ that we do not repeat it here. At $\frac{\Lambda}{T}=(\frac{\Lambda}{T})_*$, $S_{A(N)}=S_{A(C)}$. Although the number of degrees of freedom is not equal in the non-conformal and conformal field theories, the above equality indicates that  $\frac{\Lambda}{T}$ would compensate for the difference in the number of degrees of freedom. For larger values of  $\frac{\Lambda}{T}$, $\Delta S$ becomes positive and hence $S_{A(N)}>S_{A(C)}$. Note that, on the gravity side, in the low temperature limit, $\frac{\Lambda}{T}\rightarrow\infty$, the geometry approaches $AdS_5$ with radius $R_{IR}$. It seems that for large enough values of $\frac{\Lambda}{T}$, $S_{A(N)}$ does not vary significantly. We also observe that by increasing $\Lambda$, $(\frac{\Lambda}{T})_*$ decreases. Therefore, $\Delta S$ is a function of $\Lambda$ and $T$ and not only $\frac{\Lambda}{T}$. 

In figure \ref{differenM} we have plotted the $\frac{\Delta S}{S_{A(c)}}$ in terms of $\frac{\Lambda}{T}$ for two different values of $\phi_M$. Since a sizeable non-conformality emerges in the field theory for larger values of $\phi_M$, $\Delta S$ is substantially smaller for $\phi_M=1$ and large enough values of $\frac{\Lambda}{T}$.  When $\frac{\Lambda}{T}$ is small enough, the field theory is almost conformal for both cases and therefore $\Delta S$'s, for $\phi_M=1$ and $\phi_M=100$, are equal. This figure also indicates that $(\frac{\Lambda}{T})_*^{\phi_M=1}>(\frac{\Lambda}{T})_*^{\phi_M=100}$.  Since in our calculation $\Lambda$ is fixed, one can conclude that $T_*^{\phi_M=1}<T_*^{\phi_M=100}$. In other words, for $\phi_M=100$, the effect of non-conformality appears at higher temperature which is reasonable.

Another point we would like to make here is the relation between time orderings presented in \cite{Attems:2016ugt} and our observation of changing sign of $\Delta S$. At high temperatures it is shown that $t_{iso}<t_{eq}$ and in the same limit we see $\Delta S<0$. In the opposite limit, the situation is reversed, i.e. $t_{iso}>t_{eq}$ and $\Delta S>0$. Based on these observations, one may speculate that there is a connection between time ordering and the sign of $\Delta S$. This idea needs further investigation which we postpone to a future work. 

As a final point, it is instructive to investigate the possibility of a transition due to energy scale $\Lambda$. In order to do so, we have to consider two types of surfaces: connected and piecewise smooth. The connected surface has been already obtained in \eqref{entanglement} for an arbitrary background \eqref{metric1}. The second surface is defined as 
\be %
 x=-\frac{l}{2},\ \ \ \ \phi=\phi_H,\ \ \  x=\frac{l}{2},
\ee %
and then, using \eqref{area}, it is easy to find
\begin{align}\label{ss}
\hat{S}_A=\frac{V_2}{4G_5}\left(2\int_{0}^{\phi_H} d\phi f_3(\phi)\sqrt{f_2(\phi)}+l\sqrt{f_3^3(\phi_H)}\right).
\end{align}
Now let us consider the following expression 
\be %
 \bar{\Delta} S=S_A-\hat{S}_A.
\ee %
Computing difference in entanglement entropies in confining theories with gravity duals reveals a deconfinement transition \cite{Klebanov:2007ws}. An extension of above idea to the thermal backgrounds shows that entanglement entropy does not exhibit a phase transition (expect for geometry of the near horizon limit of D6-branes) \cite{Faraggi:2007fu}. In our case, transition does not take place at finite temperature as well as zero temperature. In other words, $\bar{\Delta} S$ is always negative, i.e. $S_A<\hat{S}_A$, and as a result the connected surface is favourable at all times.

\section*{Acknowledgement}
We would like to thank School of Physics of Institute
for research in fundamental sciences (IPM) for the research facilities and
environment. The authors would also like to thank H. Ebrahim, M. R. Mohammadi Mozaffar and M. M. sheikh Jabbari for useful
comments.

\end{document}